\documentclass[a4paper,11pt]{extarticle}  

\usepackage{stylesheet}

\usepackage[left=1in,top=1in,bottom=.65in,right=1in]{geometry}

\usepackage{graphicx}
\usepackage{amsfonts}
\usepackage{amsmath}
\usepackage{bm}

\title{Detection of critical events in renewable energy production time series}

\author{%
\textbf{%
Laurens Stoop \orcidlink{0000-0003-2756-5653},\textcolor{Accent}{\textsuperscript{1,2,3,*}} %
Erik Duijm \orcidlink{0000-0002-8780-7004},\textcolor{Accent}{\textsuperscript{1}} %
Ad J. Feelders \orcidlink{0000-0003-4525-1949},\textcolor{Accent}{\textsuperscript{1}} %
Machteld van den Broek\orcidlink{0000-0003-1028-1742}\textcolor{Accent}{\textsuperscript{4}}
}\\[0.5em]
\begin{small}%
\textcolor{Accent}{\textsuperscript{1}}Information and Computing Science, Utrecht University, the Netherlands \\ 
\textcolor{Accent}{\textsuperscript{2}}Copernicus Institute of Sustainable Development, Utrecht University, the Netherlands\\ 
\textcolor{Accent}{\textsuperscript{3}}TenneT TSO B.V., Arnhem, the Netherlands\\ 
\textcolor{Accent}{\textsuperscript{4}}Delft University of Technology, Faculty Technology, Policy, and Management, the Netherlands\\[0.5em] 
\textcolor{Accent}{\textsuperscript{*}}Corresponding Author: \textcolor{Accent}{laurens.stoop@tennet.eu} \\ \end{small}
}

\date{}

\begin{document}

\thispagestyle{empty}

\begin{center}

\parbox[c]{420pt}{This is the author accepted manuscript (AAM) of the following published article:}  \vspace{20pt}

\begin{tabular}{|r|l|}\hline
\textbf{DOI} & \url{https://doi.org/10.1007/978-3-030-91445-5\_7} \\ \hline
\textbf{arXiv DOI} & \url{https://doi.org/10.48550/arXiv.2401.17814} \\ \hline
 & \\ \hline
\textbf{author(s)} & \parbox{290pt}{Stoop, L.P., Duijm, E., Feelders, A., Broek, M.v.d.} \\ \hline 
\textbf{title} & \parbox{290pt}{Detection of Critical Events in Renewable Energy Production Time Series} \\ \hline
\textbf{publication date} & 01 January 2022 \\ \hline
\textbf{book} & Lecture Notes in Computer Science \\ \hline
\textbf{series} & \parbox{290pt}{Advanced Analytics and Learning on Temporal Data (AALTD) 2021} \\ \hline
\textbf{volume} & 13114 \\ \hline
\textbf{page numbers} & 104--119 \\ \hline
\end{tabular}

\vspace{30pt}
\parbox[c]{420pt}{This AAM version corresponds to the author's final version of the article, as accepted by the journal. However, it has not been copy-edited or formatted by the journal. \\[20pt]
This AAM is deposited under a Creative Commons Attribution (CC-BY-SA) license.} 
\end{center}

\newpage

\pagenumbering{arabic}

\maketitle

\begin{abstract}
The introduction of more renewable energy sources into the energy system increases the variability and weather dependence of electricity generation.
Power system simulations are used to assess the adequacy and reliability of the electricity grid over decades, but often become computational intractable for such long simulation periods with high technical detail.
To alleviate this computational burden, we investigate the use of outlier detection algorithms to find periods of extreme renewable energy generation which enables detailed modelling of the performance of power systems under these circumstances.
Specifically, we apply the Maximum Divergent Intervals (MDI) algorithm to power generation time series that have been derived from ERA5 historical climate reanalysis covering the period from 1950 through 2019.
By applying the MDI algorithm on these time series, we identified intervals of extreme low and high energy production.
To determine the outlierness of an interval different divergence measures can be used.
Where the cross-entropy measure results in shorter and strongly peaking outliers, the unbiased Kullback-Leibler divergence tends to detect longer and more persistent intervals.
These intervals are regarded as potential risks for the electricity grid by domain experts, showcasing the capability of the MDI algorithm to detect critical events in these time series.
For the historical period analysed, we found no trend in outlier intensity, or shift and lengthening of the outliers that could be attributed to climate change.
By applying MDI on climate model output, power system modellers can investigate the adequacy and possible changes of risk for the current and future electricity grid under a wider range of scenarios.
\end{abstract}

\vspace{1pc}
\noindent{\it \color{Highlight} Keywords}: Energy climate, Power system modelling, Outlier detection, Time series, Climate change, Anomaly detection, High impact events
\vspace{1pc}


\section{Introduction}
With the energy transition from fossil-fuel driven generation towards intermittent renewable energy sources like wind and solar power, the electricity supply becomes more variable\autocite{staffell2018increasing}.
Additionally, electrification of space heating will enhance\autocite{zeyringer2018designing,staffell2018increasing} the already existing variability at the electricity demand side\autocite{thornton2016role,bessec2008non}.
This twofold increase in variability can be partly counteracted by the high interconnectivity of the European electricity system\autocite{tyndp2020} that enables exchanges between countries with either electricity shortfalls or surpluses. 
However, large scale penetration of variable renewable energy sources can endanger the reliability of the system as  weather driven critical conditions may damage elements in the electricity grid or lead to hours with unserved energy\autocite{vanderwiel2019extreme}.

Therefore, insights into critical events are required to support stakeholders with taking appropriate risk reducing investments during the energy transition\autocite{zscheischlerrisk}. 
For instance, some of these critical events could be avoided by investments in flexibility options\autocite{frew2016}, interconnections\autocite{schlachtberger2017}, storage facilities\autocite{kies2016}, spatial balancing\autocite{neubacher2020multi,Grams2017weather} and/or back-up systems.

Power system simulation models can be used to select and quantify these type of investments to deal with critical events in different scenarios of power system development\autocite{Harang2020}. 
The simulations often search for cost-effective solutions under pre-set reliability and environmental performance standards. 
However, when all important features and limitations of the power system are taken into account, these power system simulations can become very complex, resulting in high computational burdens that scale with the simulation period\autocite{wuijts2022modelchar}.

These constraints on the simulation period impede that power system modellers sufficiently assess the impact of variability of intermittent renewables over different timescales ranging from sub-hourly to decadal\autocite{McCollum2020}. 
A promising method to comprehensively incorporate the variability of renewables into power system simulations without increasing the simulation period is the Importance Subsampling approach developed by \textcite{hilbers2019importance}. 
However, this method may overlook important weather-related outliers resulting in an inaccurate assessment of reliability under critical conditions. 
Although energy system experts could complement the Importance Subsampling approach with information of extreme events in the past\autocite{dawkins2020characterising}, such information is lacking for future weather years from climate models\autocite{Bloomfield2021nextgen}. 
The latter is crucial though, among others for evaluating the power system performance under climate change conditions.

In this paper we apply the Maximally Divergent Intervals (MDI) algorithm developed by \textcite{barz2017maximally} that enables the systematic detection of outliers in energy climate datasets, like renewable energy production time series.
We perform several experiments on a energy climate dataset of 70 years to determine the merits and limitations of this method to find critical events.
The developed method is a key step in a joint project with experts from a national meteorological institute and a Transmission System Operator (TSO).
It will be applied to identify critical conditions in very large datasets from climate models to assess system adequacy in many scenarios with power system simulation modelling.

This paper is organized as follows. 
Related work is discussed in Section~\ref{sec:RelatedWork} to place the outlier detection method in a broader context. 
Section~\ref{sec:data} introduces the energy climate dataset used in this study. 
Next the relevant components of the algorithm are briefly described in Section~\ref{sec:MDI}. 
The application of the algorithm is experimentally evaluated and discussed in Section~\ref{sec:experiments}. 
Finally, Section~\ref{sec:Conclusion} presents the conclusion and next steps in our project.


\section{Related work}\label{sec:RelatedWork}
Here we will focus on related work concerned with finding critical events in energy production data and weather data.
For related work on algorithms for outlier detection, we refer to the overview in the introduction of \textcite{barz2018detecting}.

A broad research community addressed the identification of extreme weather events in historical weather years by applying a variation of methods.
Where \textcite{Wu2007} focus on using Extreme Value Theory to detect and track heavy rainfall events, others like \textcite{duggimpudi2019spatio} used Behavioural outlier Factors to track the path of hurricane Katrina.

Although such extreme weather events may be of interest in their own right due to their potential severity\autocite{AR5WG2}, not all high impact events are caused by extreme weather events\autocite{vanderwiel2020}.
Therefore, research is shifting from the identification of extreme weather events to the identification of weather events that have a severe impact on society\autocite{zscheischlerrisk}.

The impact based approach asks for a clear definition of a variable that can measure the severity of this impact. 
Thus searching for weather events that pose a risk for the operation of the power system requires first of all knowledge of how weather influences the power system, secondly a method to classify the weather driven impacts between normal to adverse to severe, and thirdly to detect these severe events. 
\textcite{dawkins2020characterising}, for example created a composite impact variable capturing the relations between wind droughts and electricity demand peaks. 
Another example, is the study by \textcite{vanderwiel2019extreme} who also used a composite variable representing weather dependent solar and wind supply minus the electricity demand. 
By dividing the renewable generation by the electricity demand, significantly different critical events where found by \textcite{drew2019}, indicating the importance of the exact definition of the impact variable.

In most of these studies, the impacts are considered severe when the impact value exceeds a pre-defined threshold\autocite{dawkins2020characterising,drew2019,vanderwiel2019extreme}. 
Thus the nature of the impact is pre-determined by this selection criterion and can for example be a shortage or surplus of energy during a specific time horizon. 
Furthermore, although most studies look at extremes at different time horizons e.g., 1 day, 1 week or 2 weeks, they often fix the length of the time horizon before determining the outliers. As the intensity, duration and/or timing of high impacts can change due to climate change, a more flexible method would be beneficial when looking at climate change related risks.

Finding critical events for the power system thus requires knowledge of the relation between weather and impact. 
Expert opinion is a way to determine if an event is critical, but it might be very subjective. 
A thorough overview of critical events for the United Kingdom is given by \textcite{dawkins2020characterising} where they rely on extensive expert knowledge, and by \textcite{ward2013} for the wider region of Europe, though their work could be considered dated given the fast transition.

Additionally, despite the effort of the energy climate community the input data for such studies are not available in a coordinated way for most countries\autocite{Bloomfield2021nextgen}.
Using labelled real world data for training an outlier detection method is thus not a viable option, synthetic time series are therefore used within the energy climate community. 
This limited availability of data is especially an issue with respect to energy consumption data. 
Methods exist to model the energy consumption\autocite{moral2005modelling,thornton2016role,cassarino2018impact}, but the difficulty in the acquisition of the data required limits the scope of this paper to renewable energy generation.


\section{The Energy-Meteorological Dataset}\label{sec:data}
In this section we provide a brief introduction into the data used for our experiments and how it was generated. 
We first discuss the properties of the ERA5 dataset in Section~\ref{sec:ERA5}. 
After this we will discuss, in Section~\ref{sec:conversionmodel}, the energy conversion models used to create electricity generation data based on the ERA5 reanalysis data.

\subsection{The ERA5 Reanalysis Data}\label{sec:ERA5}
ERA5 is the latest reanalysis dataset developed by the European Centre for Medium-Range Weather Forecasts\autocite{Hersbach2020}. 
In a reanalysis dataset\autocite{ERA5}, historical observations are consistently assimilated into numerical weather models to give a best estimate of the recent climate.

ERA5 reanalysis data stretches from 1950 to the present, with a two month delay. 
The period between 1950 and 1979 is the preliminary version of the ERA5 back-extension\autocite{ERA5}. 
The ERA5 and its back-extension have undergone significant quality control and are considered state-of-the-art. 
The variables used in this research are solar irradiance, wind speed at 100 meter height, and 2 meter temperature.

The temporal granularity of the data is hourly, with a spatial granularity of 0.25 degree or $\pm$ 31 kilometers. 
The period we covered spans from 1950 through 2019. 
In the spatial domain we used the subregion of Europe, defined here as the region between latitude $-14.75$ to 40 East and longitude 35 to 74.75 North.

\subsection{Energy Conversion Models}\label{sec:conversionmodel}
To calculate the electricity generation based on climate reanalysis data one needs to know the capacity factor of wind turbines and solar photo-voltaic panels per grid cell, and the distribution of their capacity over the region of interest. 
The first can be obtained by using conversion models that compute a capacity factor for each grid cell based on the climate variables in that grid cell. 
The second, a distribution of renewable energy sources for the target year of 2050 was provided to us upon request by Bas van Zuijlen, for details on the properties of this possible distribution we refer to \textcite{VanZuijlen2019}.
\smallskip

In collaboration with the TSO stakeholder of our project, several conversion models were compared and analysed. 
For solar panel electricity generation we compared the methods presented by \textcite{Jerez2015model} and \textcite{Bett2016}. 
More advanced methods where not used as those require additional information on panel tilt, angle and solar radiation components that are not available. 
We selected the method as set out by \textcite{Jerez2015model}, we refer to them for more details.

For wind turbine electricity generation we compared the methods described by \textcite{Jerez2015model,Saint-Drenan2020,Carrillo2013,GonzalezAparicio2016,Ruiz2019}. 
Based on the model complexity, running time and accuracy of the output, we selected the general power curve method from \textcite{Jerez2015model}. 
However, we made three adjustments to this model. 
First, we reduced the effective capacity factor ($CF_e$) with $5\%$ to $95\%$ to represent the wake losses in large scale wind-farms. 
Secondly, we introduce a linear decay in the capacity factor at high wind speeds to more accurately represent high windspeed operational conditions. 
The third change was that we tuned the power curve regimes. Equation~\eqref{windpotential} gives the capacity factor for wind turbines ($CF_{wind}$) used in this study.
\begin{align}
        CF_{wind}(t) &= CF_e \times
        \begin{cases}
                0 & \mbox{if} \qquad V(t)<V_{CI},\\
                \frac{V(t)^3 - V_{CI}^3}{V_R^3-V_{CI}^3} & \mbox{if} \qquad V_{CI}\leq V(t)<V_R,\\
                1  & \mbox{if} \qquad V_R \leq V(t)<V_{D},\\
                \frac{V_{CO} - V(t)}{V_{CO}-V_{D}}  & \mbox{if} \qquad V_{D}\leq V(t)<V_{CO},\\
                0 & \mbox{if} \qquad V(t)\geq V_{CO}.
        \end{cases} \label{windpotential}
\end{align}
Here $V(t)$ is the wind speed at the height of the wind turbine and the power curve regimes are given by the cut-in ($V_{CI}$=3 m/s), rated ($V_{R}$= 11 m/s), decay ($V_{D}$= 20m/s) and cut-out ($V_{CO}$= 25m/s) wind speed. 
The windspeed provided by ERA5 (at 100 meter) did not match the hub height for offshore turbines within the capacity distribution used by \textcite{VanZuijlen2019}, therefore it is scaled using the wind profile power law to 150 meters. 
The surface roughness was set to a constant value for both onshore ($\alpha=0.143$) and offshore regions ($\alpha=0.11$).
Further details in SI Section~\ref{SIA:ECM}.

The total energy generation per grid cell is obtained by multiplying the capacity factor with the installed capacity from the distribution used.

The temporal variations in supply are expected to play a larger role than the spatial variation for the critical conditions in the power system. 
Additionally, the current European electricity grid is highly interconnected\footnote{See \url{https://www.entsoe.eu/data/map/} for a interactive map of the current network.}, even higher interconnectivity of the system is expected by 2050. 
As we search for critical conditions and we have to reduce the dataset size for tractability, we assume that the electricity grid can be approximated by a copper plate\textcite{VanZuijlen2019}. 
This implies that the flow of electricity is not impeded and local imbalances are dealt with on system wide scale.

Due to the copperplate assumption we can sum the electricity generation per technology over the European region to obtain time series data. 
Our final input time series data thus contains three variables, namely wind-onshore (WON), wind-offshore (WOF), and solar photo-voltaic (SPV) electricity generation. 
This data is based on historical weather years (1950$--$2019), but uses a possible distribution of renewables in a deep decarbonised future. 
For each variable the length of the time series is therefore $N=613,594$.

\section{The Maximally Divergent Intervals algorithm}\label{sec:MDI}
In this section we give a short description of the Maximally Divergent Intervals (MDI) algorithm (see \textcite{barz2017maximally,barz2018detecting} for more details).
This algorithm finds outliers in spatial-temporal data, but since we aggregate over the spatial component, we will restrict the presentation to the strictly temporal case.
Let
\begin{align}
\{(x_{t,1},x_{t,2},\ldots,x_{t,d}) : t &= 1,\ldots,N\}
\end{align}
be a $d$-dimensional multivariate time series of length $N$. 
Individual samples are written as $\mathbf x_t \in \mathbb R^d$.
Loosely speaking, an outlier is an interval in which the distribution of the variables deviates strongly from their distribution outside that interval. 
To model the probability distribution, Kernel Density Estimation (KDE) using Gaussian kernels or a multivariate Gaussian distribution are applied.
The anomaly score of interval $I$ is defined as:
\begin{align}
        S(I) &= \mathcal{D}(\hat{p}_I,\hat{p}_{\Omega}),\hspace{1cm}I \in \mathcal{I}
\end{align}
where $\mathcal{D}$ is some measure of the divergence between two probability distributions, $\hat{p}_I$ is the distribution fitted to the observations inside the interval, and $\hat{p}_{\Omega}$ is the distribution fitted to the remaining observations.
The set $\mathcal{I}$ contains all intervals with a time horizon length between a user-specified minimum $a$ and maximum $b$, hence $\lvert \mathcal{I} \rvert \approx N (b-a)$.
Possible divergence measures are cross-entropy, (unbiased) Kullback-Leibler and Jensen-Shannon divergence.
Although Jensen-Shannon divergence has its merits, it was found not to be tractable due to the size of our data.
The cross-entropy and Kullback-Leibler divergence are respectively computed by:
\begin{align}
\mathcal{D}_{\mbox{\sc ce}}(I,\Omega) &= \frac{1}{|I|} \sum_{t \in I} \log \hat{p}_{\Omega}(\mathbf x_t), \\
&\mbox{ and } \\
\mathcal{D}_{\mbox{\sc kl}}(I,\Omega) &= \frac{1}{|I|} \sum_{t \in I} \log \left(\frac{\hat{p}_{I}(\mathbf x_t)}{\hat{p}_{\Omega}(\mathbf x_t)}\right),
\end{align}
where $\hat{p}_{I}(\mathbf x_t)$ is the probability density of data point $t$ according to the probability density fitted to the interval, and likewise $\hat{p}_{\Omega}(\mathbf x_t)$ is the probability density of data point $t$ according to the probability density fitted to the remainder of the data. 
\textcite{barz2018detecting} note that $\mathcal{D}_{\mbox{\sc kl}}$ has a bias towards smaller intervals, and propose the unbiased variant $\mathcal{D}_{\mbox{\sc u-kl}}=2 \cdot |I| \cdot \mathcal{D}_{\mbox{\sc kl}}$.
If a multivariate Gaussian distribution is used to estimate the probability densities, then the (unbiased) Kullback-Leibler divergence can be computed quite efficiently, since in that case a closed-form solution is available.

To take into account the temporal correlation between data points, a technique call time-delay embedding is applied.
Time-delay embedding incorporates context from previous time-steps into each sample by transforming a given time series
$\{\mathbf x_t\}_{t=1}^N, \mathbf x_t \in \mathbb R^d$ into another time-series
$\{\mathbf x_t'\}_{t=1+(\kappa-1)\tau}^N$, $\mathbf x_t' \in \mathbb R^{\kappa d}$, with
\begin{align}
\mathbf x_t' &= \left(\mathbf x_t^{\top} \;\; \mathbf x_{t-\tau}^{\top} \;\; \mathbf x_{t-2\tau}^{\top} \;\; \cdots \;\; \mathbf x_{t-(\kappa-1)\tau}^{\top}\right)^{\top}.
\end{align}
Here the embedding dimension $\kappa$ specifies the number of samples to stack together and the time lag $\tau$ specifies the gap between two consecutive time-steps to be included as context.

To make the algorithm better suited for large data sets, a method that proposes intervals that are likely to contain outliers is used.
The idea behind the method is that an outlier interval tends to contain several data points that would receive high scores when using point wise outlier detection. 
One such point wise scoring method is Hotelling's $T^2$ score\textcite{hotelling1992generalization}
(or squared Mahalanobis distance):
\begin{align}
T^2_t &= (\mathbf x_t - \hat{\bm \mu})^{\top} \hat{\bm \Sigma}^{-1}(\mathbf x_t - \hat{\bm \mu}).
\end{align}
At the start and end of an outlying interval, respectively, an increase and decrease of the point wise scores is expected. Therefore, only intervals that start and end with data points whose
\begin{align}
g(t) &= \lvert T^2_{t+1} - T^2_{t-1} \rvert
\end{align}
value surpass the threshold $\theta_g = \hat{\mu}_g + \vartheta \cdot \hat{\sigma}_g$ are considered, where $\vartheta$ is a parameter to be set by the user.
Thus much less intervals need to be checked leading to a substantial speed up, since estimating distributions and divergence calculations are very time consuming.
The potential downside of this approach is that outlier intervals may be overlooked, thus lowering recall. 
However, experiments performed by \textcite{barz2018detecting} show that this was not the case when a reasonable value for $\vartheta$ was selected.

In order to ensure that the top detected outliers aren't all small variations of the same event, starting with the top outlier, the overlap:
\begin{align}
O(I_1,I_2) &= \frac{|I_1 \cap I_2|}{|I_1 \cup I_2|}
\end{align}
with lower scoring outliers is checked.
If this overlap is larger than a userdefined threshold $\theta_o$, only the interval with the higher score is reported.
Finally, the algorithm sorts the intervals in descending order of their score, so that a user-specified number of top $k$ intervals can be selected as output.

\section{Experimental Results}\label{sec:experiments}
To determine whether the MDI algorithm is suited to identify critical events in energy climate data we performed several experiments.
Each experiment represents a potential use case for our project and partners, while they are also a test case for the tuning and pre-processing used. 
The outliers found where presented to subject matter experts to determine if they provide insight in critical events that could influence the future energy system.

All experiments are performed on an Intel Xeon Gold 6130 CPU with 16 dual-cores at 2.1 GHz clock speed. 
Our setup has 125.6GB of available RAM memory. 
The multi-threading was limited to using 30 threads.

The rest of this section is organised as follows. 
First we discuss the tuning performed to make the MDI algorithm usable for renewable energy generation time series data in section~\ref{sec:tuning}. 
The top 20 outliers detected using Cross Entropy and the unbiased Kullback-Leibler divergence are then investigated in section~\ref{sec:divergence}. 
Finally in section~\ref{sec:climatechange}, we investigate if there are climate change induced changes in the intensity, time of the year and length of the top 50 outliers per decade.

\subsection{Tuning the MDI algorithm}\label{sec:tuning}
The settings of the algorithm were chosen in consultation with the domain experts, the model choices presented below are the end result.

Because the single Gaussian distribution gave quite a bad fit (see SI Section~\ref{SIB:distr}), we selected KDE using Gaussian kernels (with kernel width 1) to estimate the probability distributions.
Hotelling's $T^2$ proposal method is used with $\vartheta=1.5$. The allowed overlap between intervals was set to $\theta_o=0.5$.
The built-in data normalization method of the MDI algorithm, subtraction of the mean and division by the maximum, was used.
We used both Cross Entropy and the unbiased Kullback-Leibler divergence to score intervals.

The interval length was set to 2 days minimum, and 10 days maximum. The reason was two-fold, the usefulness of the output as deemed by our experts and tractability of the algorithm.
At shorter timescales batteries and the shifting of demand can be utilised to mitigate the effect of an outlier.
At longer timescales (sub-)seasonal storage, like hydrodams and hydrogen, can be utilised.
However, for the period between 2 and 10 days there are multiple technologies that could be utilised, some of which are not yet fully developed.
Knowledge of outliers within this window can therefore help determine what technologies could be utilised or should be further developed.
Using a minimum interval length also makes sure there is sufficient data to reliably estimate a distribution.
\begin{figure}[t!]
        \centering
        \includegraphics[width=.9\textwidth]{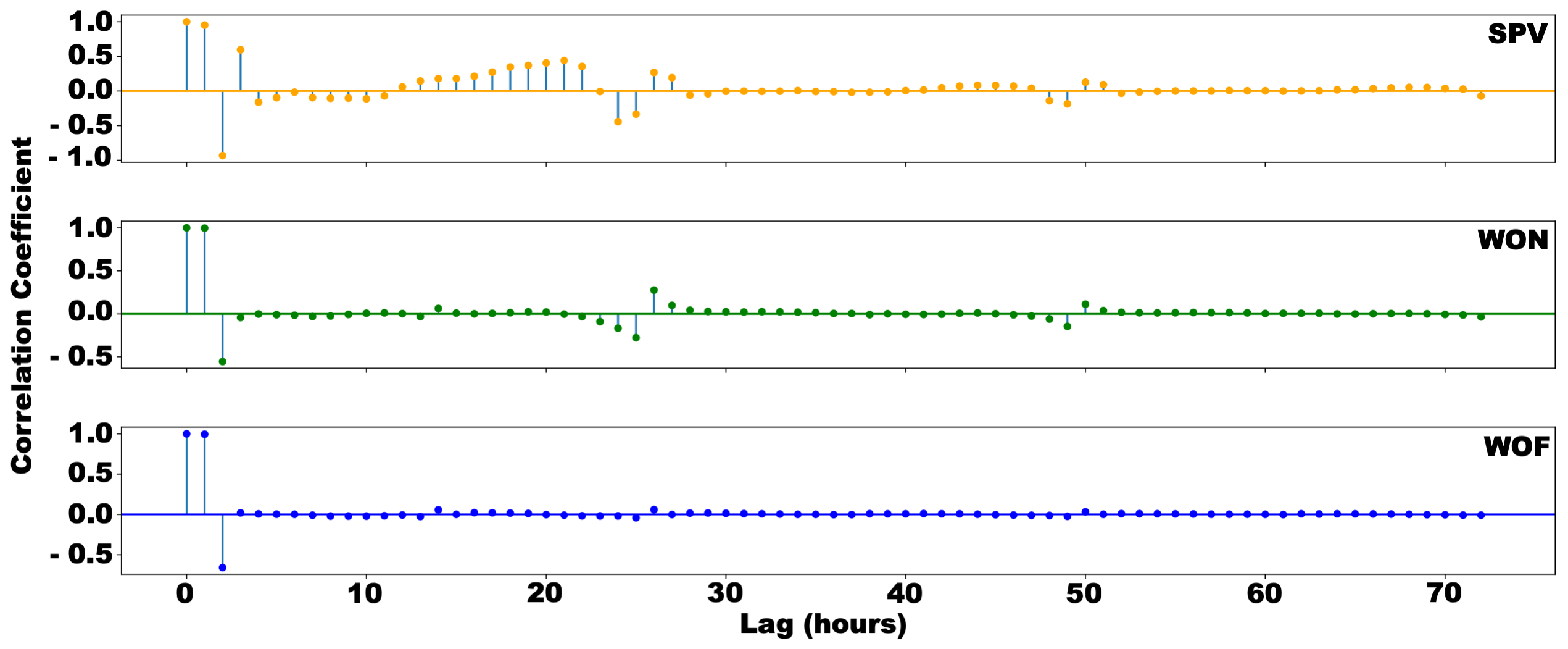}
        \caption{The autocorrelation of variables at different time lags, using the Yule-Walker method with sample size adjustment.  }
        \label{fig:pacf}
\end{figure}

In order to accurately discover temporal outliers, the temporal context embedding parameters need to be investigated. 
The idea behind the temporal context embedding is to pick points that are correlated at different time-lags. 
To investigate the autocorrelation length, the partial autocorrelation per variable was calculated (see Fig.~\ref{fig:pacf}).
Based on these calculations we have decided to use $\kappa=4$ and $\tau=8$ as respectively temporal embedding dimension and time lag settings, as these capture most of the autocorrelation in all variables. 
They ensure that the autocorrelation in solar photo-voltaic power and onshore wind power at the larger lags of approximately 24 hours are accounted for. 
These settings also ensure that at least one day and night cycle is embedded as context, which has a big impact on the Solar Photovoltaic energy generation in particular.

The original MDI algorithm of \textcite{barz2018detecting} is implemented in an open source library\footnote{\url{https://github.com/cvjena/libmaxdiv}} with both a \texttt{Python} implementation of the algorithm and a \texttt{C++} implementation. 
As the \texttt{Python} algorithm is not well suited for large data sets, we used the \texttt{C++} implementation and additionally built a wrapper in \texttt{Python} that accessed the \texttt{C++} multi-threading functionality and added \texttt{xarray} compatibility.

\subsection{Outlier Identification and Assessment}\label{sec:divergence}
\begin{landscape}
        \begin{figure}[hb]
                \centering
                \includegraphics[height=0.7\textheight, clip=true, trim=4cm 2.5cm 3.5cm 3cm]{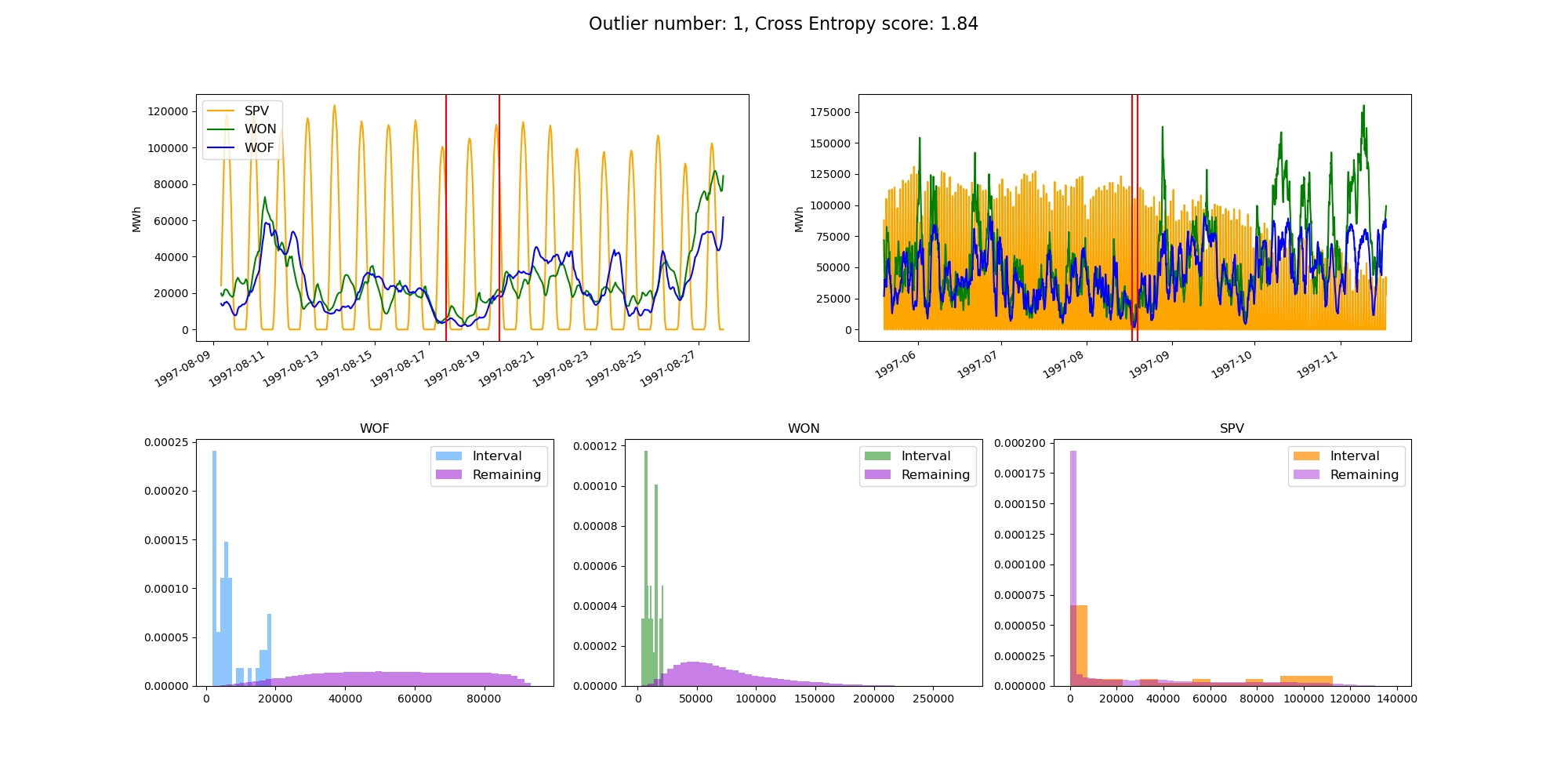}
                \caption{Figures depicting the outlier with the highest score using the Cross Entropy measure.
                        The top figures show the generation of each technology and the temporal context in which the outlier (indicated by red lines) was found.
                        The bottom images provide histograms of the generation (in MWh) of each of the three technologies during the interval (in their respective colour) and the remaining data (in purple). }
                \label{fig:Full-region-CE_norm_1}
        \end{figure}           
\end{landscape}

In order to determine if MDI can find potential shortfalls or surges that might affect the European energy system, we investigated the outliers that were identified by two divergence measures. 
The top outliers detected using Cross Entropy and the unbiased Kullback-Leibler divergence are shown in Figures~\ref{fig:Full-region-CE_norm_1} and \ref{fig:Full-region-TS_norm_1}, respectively.  
The top 20 outliers were also presented to our domain experts to harness their insight in the tuning and assessment process.
Both the Cross Entropy and unbiased Kullback-Leibler methods took just over 29 hours wall clock time to calculate.

According to the domain experts, the top 20 outliers found are all likely to be high impact events. 
Additional investigation revealed that the top outlier based on Cross Entropy coincides with a period that was identified by \textcite{dawkins2020characterising} as an adverse weather system for the electricity system of the United Kingdom and Europe. 
For the top outlier detected using the unbiased Kullback-Leibler divergence, a historical high impact event was found in the Burns' day storm (25\textsuperscript{th} January 1991). 
This storm is considered one of the worst storms of the last century for the United Kingdom, the Netherlands, and Belgium in which 97 people lost their lives.

To summarize them, the top 20 outliers were grouped based on the month in which they occur, the length of the outlier and their type. 
The type of an outlier is based on three indicators, namely Peak, Trough and Peak-Trough (see Table~\ref{tab:classCE}, \ref{tab:classuBK}). 
During a Peak, the power generation is above normal for two or more of the three energy sources. 
In a Trough, power generation is below normal. 
The Peak-Trough type indicates that the outlier contains a variable that has a peak as well as one that has a trough, and the combined energy generation over the period is neither very high nor very low.

\begin{table}[b!]
\caption{
        Grouping of the top 20 outliers found by the MDI algorithm in our time series data using Cross Entropy method. 
        The grouping has been ordered in such a way that the different outlier classes can be discerned easily. 
        It should be noted that although the length of the intervals is near the bounds, they are not at the bounds in general.}
\label{tab:classCE}
\makebox[\textwidth][c]{
\begin{tabular}{|l|c|r||c|c|c|c|}
\hline
Top-k   	&		 Month 		&	 Length(h) 	&		SPV		&		WON		&		WOF		&	Type	\\ \hline \hline
1/6/13/19	&	$	Aug.	$	&	 48-72 	&	$	+	$	&	$	-	$	&	$	-	$	&	T 	\\ \hline
3/5	&	$	June	$	&	 48-72 	&	$	+	$	&	$	-	$	&	$	-	$	&	T		\\ \hline
07/09/2017	&	$	July	$	&	 48-72 	&	$	+	$	&	$	-	$	&	$	-	$	&	T		\\ \hline
16	&	$	July	$	&	 72-96 	&	$	+	$	&	$	-	$	&	$	-	$	&	T		\\ \hline
10/15	&	$	July	$	&	 150-175 	&	$	+	$	&	$	-	$	&	$	-	$	&	T		\\ \hline
14	&	$	Feb.	$	&	 48-72 	&	$	-	$	&	$	+	$	&	$	-	$	&	PT		\\ \hline
4	&	$	Apr.	$	&	 48-72 	&	$	0	$	&	$	+	$	&	$	+	$	&	P		\\ \hline
2/11	&	$	Dec.	$	&	 48-72 	&	$	-	$	&	$	+	$	&	$	+	$	&	P		\\ \hline
12/18	&	$	Feb.	$	&	 48-72 	&	$	-	$	&	$	+	$	&	$	+	$	&	P		\\ \hline
20	&	$	Jan.	$	&	 48-72 	&	$	-	$	&	$	+	$	&	$	+	$	&	P		\\ \hline
\end{tabular} }
\end{table}

\begin{table}[b!]
\caption{
        Grouping of the top 20 outliers found by the MDI algorithm in our time series data using unbiased Kullback-Leibler method. 
        The grouping has been ordered in such a way that the different outlier classes can be discerned easily. 
        It should be noted that although the length of the intervals is near the bounds, they are not at the bounds in general.}
\label{tab:classuBK}
\makebox[\textwidth][c]{
\begin{tabular}{|l|c|r||c|c|c|c|}
\hline
Top-k	&		 Month 		&		 Length(h) 		&		SPV		&		WON		&		WOF	&	Type		\\ \hline	\hline
1/5/7/10	&	$	Jan.	$	&	$	 216+ 	$	&	$	-	$	&	$	+	$	&	$	+	$ &	P		\\ \hline
2/8/17	&	$	Dec.	$	&	$	 216+ 	$	&	$	-	$	&	$	+	$	&	$	+ $	&	P 		\\ \hline
3/4	&	$	Feb.	$	&	$	 216+ 	$	&	$	-	$	&	$	+	$	&	$	+$	&	P		\\ \hline
11/18-20	&	$	Nov.	$	&	$	 216+ 	$	&	$	-	$	&	$	+	$	&	$	+	$&	P		\\ \hline
9	&	$	Jan.	$	&	$	 192-216 	$	&	$	-	$	&	$	+	$	&	$	+	$&	P		\\ \hline
6/13/16	&	$	Feb.	$	&	$	 216+ 	$	&	$	0	$	&	$	+	$	&	$	+	$&	P		\\ \hline
14	&	$	Aug.	$	&	$	 216+ 	$	&	$	+	$	&	$	-	$	&	$	- 	$&	T 		\\ \hline
15	&	$	July	$	&	$	 216+ 	$	&	$	+	$	&	$	-	$	&	$	- 	$&	T 		\\ \hline
\end{tabular} }
\end{table}

Based on the grouping we defined classes for the outliers. 
For the unbiased Kullback-Leibler divergence these classes are Winter Surplus and Summer Deficiency. 
We consider the outliers that show a peak in the extended winter from November through March to be part of the Winter Surplus class. 
Outlier events with a trough in overall electricity generation in the extended summer period, from May through September, are part of the Summer Deficiency class. 
For Cross Entropy we have similar classes: Winter Surplus, Long Term Summer Deficiency and Short Term Summer Deficiency. 
The only distinction is that for the Summer Deficiency we have sub classes based on the length of the event: outliers that last between 48 and 72 hours are considered short term, and outlier events longer than 72 hours are considered long term.

These classes can be problematic as they influence the whole network. 
A long deficiency needs to be compensated with other methods of non-carbon generation that need to be flexible and can be controlled, as the current battery capacities aren't sufficient. 
Shorter deficiencies during the summer are also problematic, as they require extensive use of battery capacity. 
During the day the batteries charge on the available solar photo-voltaic energy generation, but at night they need to be discharged to compensate for the lack of wind. 
This strain on the batteries causes them to wear. 
An increase in such short deficiencies represents an economic risk, as the batteries would need to be replaced more frequently. 
The Winter Surplus increases the energy generation of the grid, causing a surplus, which can be problematic if this isn't controlled. 
The surplus needs to be discharged somehow. 
This discharge of unused energy represents an economic risk, as the wind turbines and solar panels are wearing down, without the energy that is generated being used.

Based on the top 20 outliers we note that the outliers detected by the Cross Entropy measure tend to have a very short duration, whereas the outliers detected by the unbiased Kullback-Leibler divergence tend to be longer.
As a quick reminder, Cross Entropy is related to the Kullback-Leibler divergence measure and the latter was found by \textcite{barz2018detecting} to have a bias towards smaller intervals. 
We can thus expect this tendency to shorter intervals for Cross Entropy outliers. 
However, the tendency towards longer intervals is unexpected for the unbiased Kullback-Leibler divergence measure as it was created specifically to be unbiased towards interval length.
It should be noted that while some outliers are found on the bounds set on the outlier duration, they are in general not on these bounds.

Irrespective of the tendency to be near the boundary interval lengths, both divergence measure studies where deemed to identify likely high impact events by our domain experts. 
Therefore both measures should be considered when studying high impact events in energy climate data.

\subsection{Historic Climate Variability and Change}\label{sec:climatechange}
The change of intensity, time of the year, and length of outliers might change the impact of an event and is therefore important to consider\textcite{vanderwiel2020,Harang2020}.
For these experiments we combined offshore wind, onshore wind and solar photo-voltaic power generation into a single variable called Total Electricity Generation (TEG). 
This single aggregate provides a reasonable indication of shortages and surges in the electricity system, while reducing the computational burden of the algorithm. 
We identify the top 50 outliers per decade and use these in our assessment of the intensity, time and length of the outliers over the historic period.

We found that the outliers in the TEG time series represent mostly peaks.
Trough-type outliers were difficult to detect in the TEG dataset, especially when using the Cross Entropy measure. 
Potentially risky situations as in Figure~\ref{fig:Full-region-CE_norm_1} remain undetected in this univariate analysis.
This underlines the added insight provided by the multivariate analysis, and highlights the importance of selecting the correct divergence measure.
\begin{figure}[!b]
        \centering
        \includegraphics[width =\textwidth, clip=true, trim=0 0 0 0]{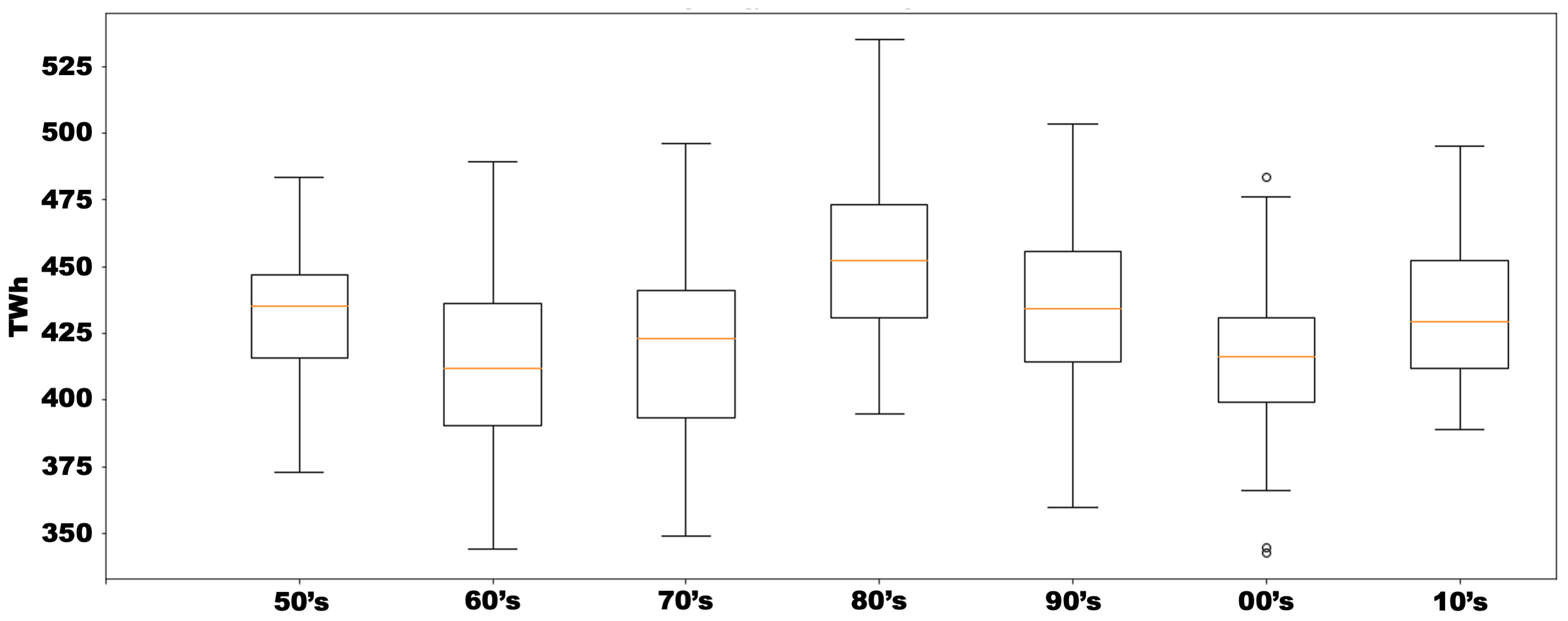}
        \caption{
                Boxplot of the average hourly Total Energy Generation during the top 50 outlier events per decade based on the Cross Entropy measure.}
        \label{fig:box_ce_50}
\end{figure}

The intensity of the outliers is investigated by looking at the average energy generation during the outlier.
Figure~\ref{fig:box_ce_50} shows a boxplot of the average Total Energy Generation during the outlier for the top 50 outliers found with Cross Entropy divergence.
While there is no linear trend visible, some periodical behaviour appears to influence the outlier events. 
This periodic behaviour appears in all combinations of top number of outliers investigated and divergence measures used. 
Due to the presence of Trough-type outlying events this effect is hard see for the unbiased Kullback-Leibler divergence (figure not shown). 
Similar behaviour of multidecadel variability in German wind energy generation was found by \textcite{wohland2019significant}.

These result emphasise that the multidecadel variability needs to be taken into account by policy makers as it influences the strength of the outliers. 
Assessments of the energy system currently only use a limited set of weather years and might therefore under- or overestimate the extremeness of critical conditions for the energy system.

We studied the timing and duration of the outliers found per decade in the TEG time series to determine whether they are affected by climate change. 
We did, however, not find any obvious trends or shifts that could potentially be attributed to climate change. 
See Section~\ref{SIC:CC} for more details.

\begin{landscape}
        \vspace*{\fill}
        \begin{figure}[hb]
                \centering
                \includegraphics[height=0.7\textheight, clip=true, trim=4cm 2.5cm 3.5cm 3cm]{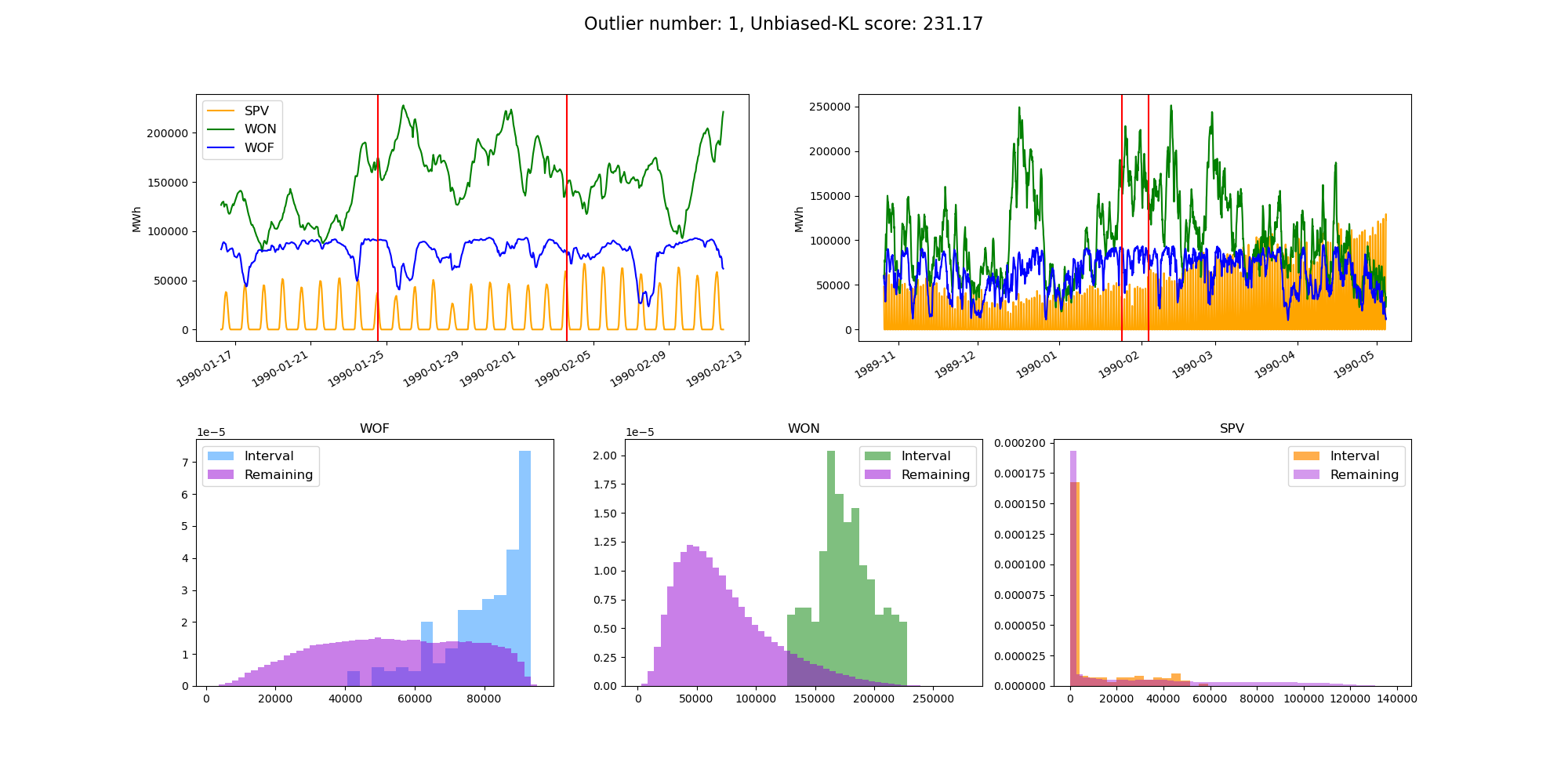}
                \caption{
                        Figures depicting the outlier with the highest score using the unbiased Kullback-Leibler divergence measure. 
                        As shown in Figure~~\ref{fig:Full-region-CE_norm_1}. }
                \label{fig:Full-region-TS_norm_1}
        \end{figure}    
        \vspace*{\fill}          
\end{landscape}

Such trends or shifts are possibly masked by the multidecadel variability in the outliers. 
The time of emergence of a climate change signal lies thus in the future, like it currently does for most climate related impacts\autocite{Hawkins2012}.

\section{Conclusion and Future work}\label{sec:Conclusion}
Using the Maximally Divergent Intervals (MDI) algorithm we found outlying time periods in 70 years of historic weather-derived energy production data on three types of renewable energy. 
According to subject area experts from a national Transmission System Operator (TSO),
the identified outliers indeed represented periods during which the European electricity system could be at risk. 
However, when the three renewable energy generation variables were combined into a single variable, Total Energy Generation, potential outliers were missed as mostly peak-type outliers were detected. 
The multivariate analysis is therefor preferred in further work.
We conclude that, with the proper parameter settings, outlier detection with MDI can help the assessment of the future energy grid by highlighting the most extreme situations.

When analysing the Total Energy Generation peer decade we found that the intensity of outliers manifests multidecadel variability over the last 70 years. 
However, we found no trend could be attributed to climate change.
This variability in the outliers also hinders the determination of climate change attributable shift or duration change in the historic period.

We demonstrated the added value of outlier detection with the MDI algorithm compared to existing methods that require an a priori specification of the critical events to be detected.
Experiments showed that both outliers of a different nature as well as with varying lengths were detected. 
Additionally, as the length of the outlier interval is not a fixed in advance, comparison between events of different lengths is possible. 
However, there is a dependency between the lengths of the detected intervals, and the divergence measure used. 
Cross Entropy tends to prefer intervals of shorter duration, while the unbiased Kullback-Leibler divergence tends to prefer longer intervals.
As both measures provide useful insights according to our subject area experts, we will continue to use both measures for outlier detection in energy climate data.
\smallskip

In the next phase of the project, the method presented here will be used for two applications related to the assessment of power system adequacy.
First, when the outliers identified are combined with
a method to represent the generic variability of the weather, a synthetic representative time series  could be constructed.
Power system simulations based on a synthetic time series can be used to ensure both representativeness with respect to critical climate conditions as well as computational tractability.
Second, besides applying the MDI method on a historical climate dataset as was demonstrated in this paper, it can be applied to climate projections from a multitude of climate simulation models to investigate how climate change and multidecadal climate variability influence the character and frequency of critical conditions for the electricity grid.

For both applications, the method is preferably applied to a dataset that also takes electricity demand into account.
For this purpose, the temperature dependant component of the electricity demand should be based on climate variables used for the calculation of the electricity generation from renewable sources. Incorporation of energy consumption data might decrease or exacerbate the impact of critical weather events.


\section*{CRediT Author Statement}
Conceptualization: Methodology: Investigation: Visualization: Writing:  Editing: Funding Acquisition: Supervision: . 

\section*{Acknowledgments}
The data used in the experiments contains modified Copernicus Climate Change Service information, \url{doi.org/10.24381/cds.adbb2d47} (2020). 
This research received funding from the Netherlands Organisation for Scientific Research (NWO) under grant number 647.003.005.
The methodology presented here was developed as part of the IS-ENES3 project that has received funding from the European Union’s Horizon 2020 research and innovation programme under grant agreement No 824084.
The content of this paper and the views expressed in it are solely the author’s responsibility, and do not necessarily reflect the views of TenneT TSO B.V..

\section*{Open research} 
The implementation of the MDI algorithm presented, the power generation data summed for the European region, and code used for the findings presented in this study are available at Github via \url{https://github.com/laurensstoop/outlier_detection} with the MIT license. 

To use our \texttt{Python} wrapper for the MDI algorithm, please use the following instructions:
\begin{enumerate}
        \item Obtain the implementation from the original authors as found on: \\ \url{https://github.com/cvjena/libmaxdiv}.
        \item Move our python wrapper to the folder \texttt{MDI/libmaxdiv/maxdiv/CodeECMLPKDD21.py}.
        \item To run the MDI algorithm, change the \texttt{CodeECMLPKDD21.py} file such that the parameters are suited for your needs, and the data paths point to the \texttt{xarray} data
        \item Then simply run the \texttt{CodeECMLPKDD21.py} file in a suitable environment
\end{enumerate}

\printbibliography

@article{AR5WG2,
    title        = {Key economic sectors and services},
    author       = {Arent, Douglas J and Tol, Richard SJ and Faust, Eberhard and Hella, Joseph P and Kumar, Surender and Strzepek, Kenneth M and T{\'o}th, Ferenc L and Yan, Denghua and Abdulla, Amjad and Kheshgi, Haroon and others},
    year         = 2015,
    booktitle    = {Climate change 2014 impacts, adaptation and vulnerability: Part a: Global and sectoral aspects}
}

@inproceedings{barz2017maximally,
    title        = {Maximally divergent intervals for extreme weather event detection},
    author       = {Barz, Bj{\"o}rn and Garcia, Yanira Guanche and Rodner, Erik and Denzler, Joachim},
    year         = 2017,
    booktitle    = {OCEANS 2017-Aberdeen},
    doi          = {10.1109/OCEANSE.2017.8084569}
}

@article{barz2018detecting,
    title        = {Detecting regions of maximal divergence for spatio-temporal anomaly detection},
    author       = {Barz, Bj{\"o}rn and Rodner, Erik and Garcia, Yanira Guanche and Denzler, Joachim},
    year         = 2018,
    journal      = {IEEE transactions on pattern analysis and machine intelligence},
    doi          = {10.1109/TPAMI.2018.2823766}
}

@article{bessec2008non,
    title        = {The non-linear link between electricity consumption and temperature in Europe: A threshold panel approach},
    author       = {Bessec, Marie and Fouquau, Julien},
    year         = 2008,
    journal      = {Energy Economics},
    doi          = {10.1016/j.eneco.2008.02.003}
}

@article{Bett2016,
    title        = {The climatological relationships between wind and solar energy supply in Britain},
    author       = {Bett, Philip E. and Thornton, Hazel E.},
    year         = 2016,
    journal      = {Renewable Energy},
    doi          = {10.1016/j.renene.2015.10.006}
}

@article{Bloomfield2021nextgen,
    title        = {The Importance of Weather and Climate to Energy Systems: A Workshop on Next Generation Challenges in Energy{\textendash}Climate Modeling},
    author       = {H. C. Bloomfield and P. L. M. Gonzalez and J. K. Lundquist and L. P. Stoop and J. Browell and R. Dargaville and M. {De Felice} and K. Gruber and A. Hilbers and A. Kies and M. Panteli and H. E. Thornton and J. Wohland and M. Zeyringer and D. J. Brayshaw},
    year         = 2021,
    journal      = {Bulletin of the American Meteorological Society},
    doi          = {10.1175/bams-d-20-0256.1}
}

@article{Carrillo2013,
    title        = {Review of power curve modelling for wind turbines},
    author       = {Carrillo, C and Monta{\~n}o, AF Obando and Cidr{\'a}s, J and D{\'\i}az-Dorado, E},
    year         = 2013,
    journal      = {Renewable and Sustainable Energy Reviews},
    doi          = {10.1016/j.rser.2013.01.012}
}

@article{cassarino2018impact,
    title        = {The impact of social and weather drivers on the historical electricity demand in Europe},
    author       = {Cassarino, Tiziano Gallo and Sharp, Ed and Barrett, Mark},
    year         = 2018,
    journal      = {Applied energy},
    doi          = {10.1016/j.apenergy.2018.07.108}
}

@misc{dawkins2020characterising,
    title        = {Characterising Adverse Weather for the UK Electricity System},
    author       = {Dawkins, Laura and Rushby, Isabel},
    year         = 2021,
    url          = {nic.org.uk/app/uploads/MetOffice-Characterising-Adverse-Weather-Phase-2a.pdf}
}

@article{drew2019,
    title        = {Sunny windy sundays},
    author       = {Daniel R. Drew and Phil J. Coker and Hannah C. Bloomfield and David J. Brayshaw and Janet F. Barlow and Andrew Richards},
    year         = 2019,
    journal      = {Renewable Energy},
    doi          = {10.1016/j.renene.2019.02.029}
}

@article{duggimpudi2019spatio,
    title        = {Spatio-temporal outlier detection algorithms based on computing behavioral outlierness factor},
    author       = {Duggimpudi, Maria Bala and Abbady, Shaaban and Chen, Jian and Raghavan, Vijay V},
    year         = 2019,
    journal      = {Data \& Knowledge Engineering},
    doi          = {10.1016/j.datak.2017.12.001}
}

@misc{ERA5,
    title        = {{ERA5} documentation, \textit{Contains modified Copernicus Climate Change Service Information 2020}},
    author       = {Copernicus Climate Change Service},
    doi          = {10.24381/cds.adbb2d47},
    url          = {https://cds.climate.copernicus-climate.eu/}
}

@article{frew2016,
    title        = {Flexibility mechanisms and pathways to a highly renewable US electricity future},
    author       = {Bethany A. Frew and Sarah Becker and Michael J. Dvorak and Gorm B. Andresen and Mark Z. Jacobson},
    year         = 2016,
    journal      = {Energy},
    doi          = {10.1016/j.energy.2016.01.079}
}

@misc{GonzalezAparicio2016,
    title        = {{EMHIRES dataset Part I: Wind power generation}},
    author       = {Gonzalez Aparicio I and Zucker A and Careri F and Monforti-Ferrario F and Huld T and Badger J},
    year         = 2016,
    booktitle    = {European Meteorological derived High resolution RES generation time series for present and future scenarios},
    doi          = {10.2790/831549},
    isbn         = {978-92-79-63193-1}
}

@article{Grams2017weather,
    title        = {Balancing Europe's wind-power output through spatial deployment informed by weather regimes},
    author       = {Christian M. Grams and Remo Beerli and Stefan Pfenninger and Iain Staffell and Heini Wernli},
    year         = 2017,
    journal      = {Nature Climate Change},
    doi          = {10.1038/nclimate3338}
}

@article{Harang2020,
    title        = {Incorporating climate change effects into the European power system adequacy assessment using a post-processing method},
    author       = {In\`{e}s Harang and Fabian Heymann and Laurens P. Stoop},
    year         = 2020,
    journal      = {Sustainable Energy, Grids and Networks},
    doi          = {10.1016/j.segan.2020.100403}
}

@article{Hawkins2012,
    title        = {Time of emergence of climate signals},
    author       = {Hawkins, E. and Sutton, R.},
    year         = 2012,
    journal      = {Geophysical Research Letters},
    doi          = {10.1029/2011GL050087}
}

@article{Hersbach2020,
    title        = {The {ERA}5 global reanalysis},
    author       = {Hans Hersbach and Bill Bell and Paul Berrisford and Shoji Hirahara and Andr{\'{a}}s Hor{\'{a}}nyi and Joaqu{\'{\i}}n Mu{\~{n}}oz-Sabater and Julien Nicolas and Carole Peubey and Raluca Radu and Dinand Schepers and Adrian Simmons and Cornel Soci and Saleh Abdalla and Xavier Abellan and Gianpaolo Balsamo and Peter Bechtold and Gionata Biavati and Jean Bidlot and Massimo Bonavita and Giovanna Chiara and Per Dahlgren and Dick Dee and Michail Diamantakis and Rossana Dragani and Johannes Flemming and Richard Forbes and Manuel Fuentes and Alan Geer and Leo Haimberger and Sean Healy and Robin J. Hogan and El{\'{\i}}as H{\'{o}}lm and Marta Janiskov{\'{a}} and Sarah Keeley and Patrick Laloyaux and Philippe Lopez and Cristina Lupu and Gabor Radnoti and Patricia Rosnay and Iryna Rozum and Freja Vamborg and Sebastien Villaume and Jean-No\"{e}l Th{\'{e}}paut},
    year         = 2020,
    journal      = {Quarterly Journal of the Royal Meteorological Society},
    doi          = {10.1002/qj.3803}
}

@article{hilbers2019importance,
    title        = {Importance subsampling: improving power system planning under climate-based uncertainty},
    author       = {Hilbers, Adriaan P and Brayshaw, David J and Gandy, Axel},
    year         = 2019,
    journal      = {Applied Energy},
    doi          = {10.1016/j.apenergy.2019.04.110}
}

@incollection{hotelling1992generalization,
    title        = {The generalization of Student's ratio},
    author       = {Hotelling, Harold},
    year         = 1992,
    booktitle    = {Breakthroughs in statistics},
    doi          = {10.1007/978-1-4612-0919-5\_4}
}

@article{Jerez2015model,
    title        = {{The CLIMIX model: A tool to create and evaluate spatially-resolved scenarios of photovoltaic and wind power development}},
    author       = {Jerez, S. and Thais, F. and Tobin, I. and Wild, M. and Colette, A. and Yiou, P. and Vautard, R.},
    year         = 2015,
    journal      = {Renewable and Sustainable Energy Reviews},
    doi          = {10.1016/j.rser.2014.09.041}
}

@article{kies2016,
    title        = {The Effect of Hydro Power on the Optimal Distribution of Wind and Solar Generation Facilities in a Simplified Highly Renewable European Power System},
    author       = {Alexander Kies and Bruno U. Schyska and Lueder {von Bremen}},
    year         = 2016,
    journal      = {Energy Procedia},
    doi          = {10.1016/j.egypro.2016.10.043}
}

@article{McCollum2020,
    title        = {Energy modellers should explore extremes more systematically in scenarios},
    author       = {McCollum, David L and Gambhir, Ajay and Rogelj, Joeri and Wilson, Charlie},
    year         = 2020,
    journal      = {Nature Energy},
    doi          = {10.1038/s41560-020-0555-3}
}

@article{moral2005modelling,
    title        = {Modelling the non-linear response of Spanish electricity demand to temperature variations},
    author       = {Moral-Carcedo, Julian and Vic{\'e}ns-Otero, Jos{\'e}},
    year         = 2005,
    journal      = {Energy economics},
    doi          = {10.1016/j.eneco.2005.01.003}
}

@article{neubacher2020multi,
    title        = {Multi-decadal offshore wind power variability can be mitigated through optimized European allocation},
    author       = {Neubacher, C. and Witthaut, D. and Wohland, J.},
    year         = 2021,
    journal      = {Advances in Geosciences},
    doi          = {10.5194/adgeo-54-205-2021}
}

@article{Ruiz2019,
    title        = {{ENSPRESO - an open, EU-28 wide, transparent and coherent database of wind, solar and biomass energy potentials}},
    author       = {Ruiz, P. and et al.},
    year         = 2019,
    journal      = {Energy Strategy Reviews},
    doi          = {10.1016/j.esr.2019.100379}
}

@article{Saint-Drenan2020,
    title        = {{A parametric model for wind turbine power curves incorporating environmental conditions}},
    author       = {Saint-Drenan, Yves Marie and {et al.}},
    year         = 2020,
    journal      = {Renewable Energy},
    doi          = {10.1016/j.renene.2020.04.123}
}

@article{schlachtberger2017,
    title        = {The benefits of cooperation in a highly renewable European electricity network},
    author       = {D.P. Schlachtberger and T. Brown and S. Schramm and M. Greiner},
    year         = 2017,
    journal      = {Energy},
    doi          = {10.1016/j.energy.2017.06.004}
}

@article{staffell2018increasing,
    title        = {The increasing impact of weather on electricity supply and demand},
    author       = {Staffell, Iain and Pfenninger, Stefan},
    year         = 2018,
    journal      = {Energy},
    doi          = {10.1016/j.energy.2017.12.051}
}

@article{thornton2016role,
    title        = {The role of temperature in the variability and extremes of electricity and gas demand in Great Britain},
    author       = {Thornton, HE and Hoskins, Brian J and Scaife, AA},
    year         = 2016,
    journal      = {Environmental Research Letters},
    doi          = {10.1088/1748-9326/11/11/114015}
}

@techreport{tyndp2020,
    title        = {Ten-year network development plan 2020},
    author       = {{ENTSO-E}},
    year         = 2021,
    url          = {https://eepublicdownloads.blob.core.windows.net/public-cdn-container/tyndp-documents/TYNDP2020/Forconsultation/TYNDP2020\_Report\_forconsultation.pdf},
    institution  = {{European Network of Transmission System Operators for Electricity, Brussels}}
}

@article{vanderwiel2019extreme,
    title        = {Meteorological conditions leading to extreme low variable renewable energy production and extreme high energy shortfall},
    author       = {K. {Van der Wiel} and L.P. {Stoop} and B.R.H. {van Zuijlen} and R. Blackport and M.A. {van den Broek} and F.M. Selten},
    year         = 2019,
    journal      = {Renewable and Sustainable Energy Reviews},
    doi          = {10.1016/j.rser.2019.04.065}
}

@article{vanderwiel2020,
    title        = {Ensemble climate-impact modelling: extreme impacts from moderate meteorological conditions},
    author       = {K. {van der Wiel} and Frank M Selten and Richard Bintanja and Russell Blackport and James A Screen},
    year         = 2020,
    journal      = {Environmental Research Letters},
    doi          = {10.1088/1748-9326/ab7668}
}

@article{VanZuijlen2019,
    title        = {{Cost-optimal reliable power generation in a deep decarbonisation future}},
    author       = {{van Zuijlen}, Bas and Zappa, William and Turkenburg, Wim and {van der Schrier}, Gerard and {van den Broek}, Machteld},
    year         = 2019,
    journal      = {Applied Energy},
    doi          = {10.1016/j.apenergy.2019.113587}
}

@article{ward2013,
    title        = {The effect of weather on grid systems and the reliability of electricity supply},
    author       = {Ward, David M},
    year         = 2013,
    journal      = {Climatic Change},
    doi          = {10.1007/s10584-013-0916-z}
}

@article{wohland2019significant,
    title        = {Significant multidecadal variability in German wind energy generation},
    author       = {Jan Wohland and Nour Eddine Omrani and Noel Keenlyside and Dirk Witthaut},
    year         = 2019,
    journal      = {Wind Energy Science},
    doi          = {10.5194/wes-4-515-2019}
}

@inproceedings{Wu2007,
    title        = {Spatio-temporal analysis of the relationship between south american precipitation extremes and the el ni{\~n}o southern oscillation},
    author       = {Wu, Elizabeth and Chawla, Sanjay},
    year         = 2007,
    booktitle    = {ICDMW 2007},
    doi          = {10.1109/ICDMW.2007.102}
}

@article{wuijts2022modelchar,
    title        = {Effect of modelling choices in the unit commitment problem},
    author       = {Rogier H. Wuijts and J.M. {van den Akker} and Machteld van den Broek},
    year         = 2023,
    journal      = {Energy Systems},
    doi          = {10.1007/s12667-023-00564-5}
}

@article{zeyringer2018designing,
    title        = {Designing low-carbon power systems for Great Britain in 2050 that are robust to the spatiotemporal and inter-annual variability of weather},
    author       = {Zeyringer, Marianne and Price, James and Fais, Birgit and Li, Pei-Hao and Sharp, Ed},
    year         = 2018,
    journal      = {Nature Energy},
    doi          = {10.1038/s41560-018-0128-x}
}

@incollection{zscheischlerrisk,
    title        = {Multivariate extremes and compound events},
    author       = {Jakob Zscheischler and Bart {van den Hurk} and Philip J. Ward and Seth Westra},
    year         = 2020,
    booktitle    = {Climate Extremes and Their Implications for Impact and Risk Assessment},
    doi          = {10.1016/B978-0-12-814895-2.00004-5}
}

\appendix
\beginsupplement

\section{Energy Converion Models}\label{SIA:ECM}
\subsection{Wind energy conversion model}
For clarity the wind energy conversion model is repeated here, with some specific details. We made two adjustments to the power curve method from~\cite{Jerez2015model} in collaboration with the TSO stakeholder of our project. We limited the effective capacity factor with $5\%$, introduced a quadratic decay in the capacity factor at high wind speeds and set the power curve regimes. Equation \eqref{windpot} gives the capacity factor for wind energy ($CF_{wind}$) used in this study.
\begin{align}
       CF_{wind}(t) &= 0.95 \times
       \begin{cases}
               0 & \mbox{if} \qquad V(t)<V_{CI},\\
               \frac{V(t)^3 - V_{CI}^3}{V_R^3-V_{CI}^3} & \mbox{if} \qquad V_{CI}\leq V(t)<V_R,\\
               1  & \mbox{if} \qquad V_R \leq V(t)<V_{D},\\
               \frac{V_{CO}^3 - V(t)^3}{V_{CO}^3-V_{D}^3} & \mbox{if} \qquad V_{D}\leq V(t)<V_{CO},\\
               0 & \mbox{if} \qquad V(t)\geq V_{CO}.
       \end{cases} \label{windpot}
\end{align}
Here $V(t)$ is the wind speed at the height of the wind turbine and the power curve regimes are given by the cut-in ($V_{CI}$=3 m/s), rated ($V_{R}$= 11 m/s), decay ($V_{D}$= 20m/s) and cut-out ($V_{CO}$= 25m/s) wind speed. If the wind speed given by the climate model is not equal to the hub height of the windturbine, the windspeed is scaled using the scaling law formula (see Eq. \eqref{scaling}).
\begin{align}
       V(h,t) &=  V(h_0,t)\left(\frac{h}{h_0} \right)^\alpha\label{scaling}
\end{align}
in which $h_0$ is the initial height and $h$ the desired height, which is 150 meters for offshore and 100 meters for onshore turbines in the capacity distribution used~\cite{VanZuijlen2019}. The scaling factor $\alpha$ is dependent on the surface roughness and set to a constant value for both onshore ($\alpha=0.143$) and offshore regions ($\alpha=0.11$).

\subsection{Solar Photovoltaic conversion model}
The conversion model used to obtain the solar photo-voltaic energy capacity factors follows the method as set out by \cite{Bett2016}. This method is used as no assumptions have to be made about the specific properties of the solar panels, and no additional meteorological variables are required.

Equation \eqref{solarpot} gives the capacity factor for solar photovoltaic panels ($CF_{PV}$) used in this study.
\begin{align}
       CF_{PV} &= \eta_{rel} \times \frac{G(t)}{G_{STC}} \label{solarpot}
\end{align}
Here $G(t)$ is the solar irradiance and the irradiation standard $G_{STC}= 1000W/m^2$. The relative efficiency $\eta_{rel}$ given by \cite{Bett2016} is not explicit, but rewriting it results in:
\begin{align}
       \eta_{rel}(t) &= \left[1+\alpha \Delta T_{mod}(t)\right] \times \left[1 + c_1 \ln\left(\frac{G(t)}{G_{STC}}\right) + c_2 \ln^2\left(\frac{G(t)}{G_{STC}}\right) + \beta \Delta T_{mod}(t) \right]
\end{align}
Here $\alpha$, $c_1$, $c_2$ and $\beta$ are constants and the emperical relation for model temperature ($\Delta T_{mod}(t)$) is given by:
\begin{align}
       \Delta T_{mod}(t) &= T(t) - T_{STC} + (T_{NOCT} - T_O) \times \frac{G(t)}{G_O},
\end{align}
where $T_{STC}$, $T_{NOCT}$, $T_{O}$ and $G_O$ are constants. All constants used in the Solar PhotoVoltaic conversion model are given in table \ref{tab:solartable}.

\begin{table}[h!]
       \caption{Value of constants used for solar conversion model, the values follow~\cite{Bett2016}.  }
       \label{tab:solartable}
       \centering
       \begin{tabular}{lr|l}
               Constant		&		Value		&		Unit		\\ \hline \hline
               $	\alpha	$	&	$	4.2\times 10^{-3}	$	&	$	\degree C^{-1}	$	\\ \hline
               $	\beta	$	&	$	-4.60\times 10^{-3}	$	&	$	\degree C^{-1}	$	\\ \hline
               $	c_{1}	$	&	$	0.033	$	&	$		$	\\ \hline
               $	c_{2}	$	&	$	-0.092	$	&	$		$	\\ \hline
               $	G_{STC}	$	&	$	1000	$	&	$	W/m^2	$	\\ \hline
               $	T_{STC}	$	&	$	25	$	&	$	\degree C	$	\\ \hline
               $	T_0	$	&	$	20	$	&	$	\degree C	$	\\ \hline
               $	G_0	$	&	$	800	$	&	$	W/m^2	$	\\ \hline
               $	T_{NOCT}	$	&	$	48	$	&	$	\degree C	$\\\hline
       \end{tabular}
\end{table}

A small adjustment in the implementation was made as rounding errors caused small ($<0.1\%$ of peak) negative values in summer twilight periods. By setting negative values to zero after the application of the conversion model this issue was solved.

\section{Selection of the distribution model}\label{SIB:distr}
\begin{figure}[ht!]
        \centering
        \includegraphics[width=\textwidth, clip=true, trim=3cm 1cm 3cm 1cm]{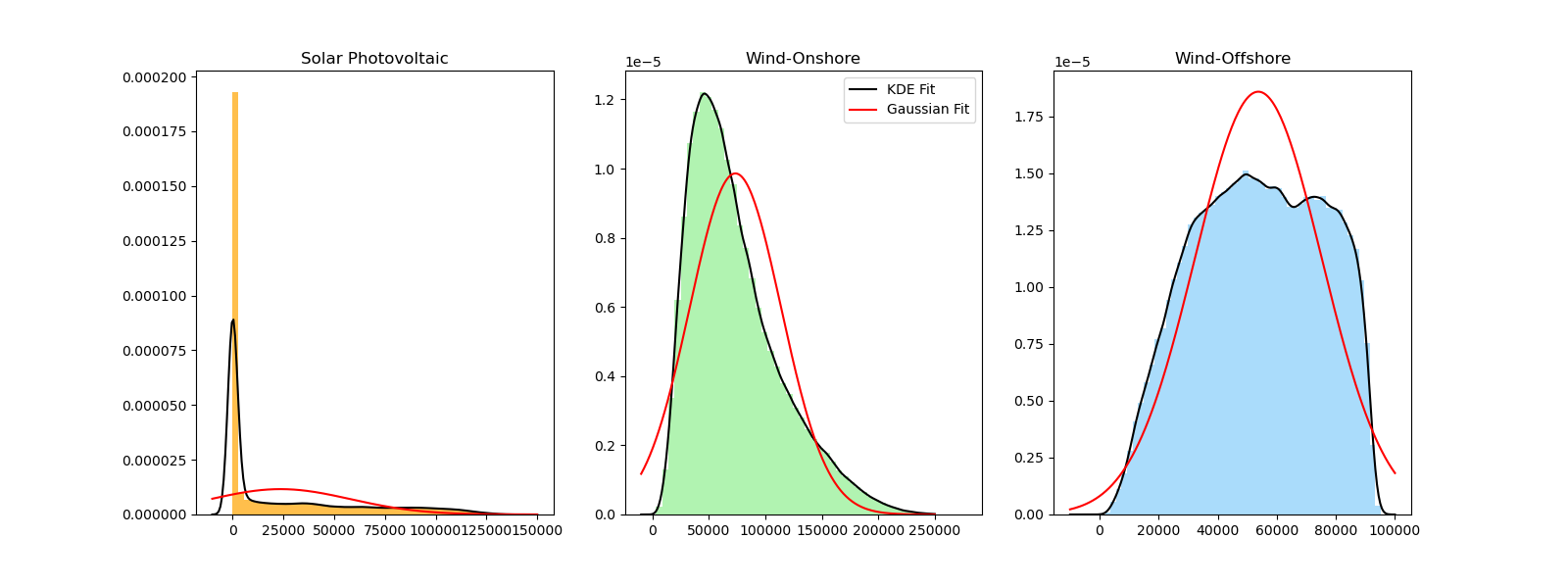}
        \caption{Histograms of the Solar Photovoltaic, Wind-Onshore and Wind-Offshore energy generation time series, plotted together with a fitted Normal Distribution and a fitted KDE using Gaussian Kernels.}
        \label{fig:total_region_fits}
\end{figure}

In order to select a suitable distribution for the data, the data was fitted to both a Gaussian and a KDE using Gaussian Kernels (with kernel width $h=1$).
These fits can be seen in figure \ref{fig:total_region_fits}, and show that the Gaussian fit is quite bad, whereas the KDE provides a good fit.
Therefore we selected the KDE for our experiments.

\section{Climate change and decadal variability}\label{SIC:CC}
In this section some additional information is provided regarding the climate change experiment that was conducted.

\subsection{Outlier Timing}
\begin{figure}[hb]
        \centering
        \includegraphics[width = \textwidth]{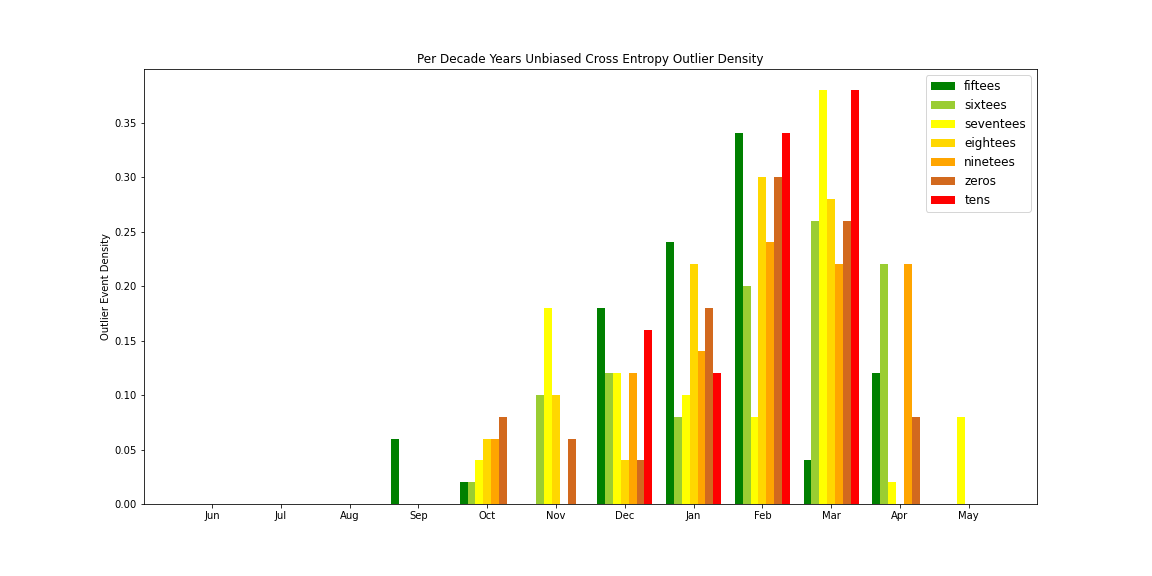}
        \caption{Histogram of the monthly occurrence of outliers per decade. Results for the Cross Entropy divergence measure.}
        \label{fig:timing_ce_decades}
\end{figure}

\begin{figure}[ht]
        \centering
        \includegraphics[width = \textwidth]{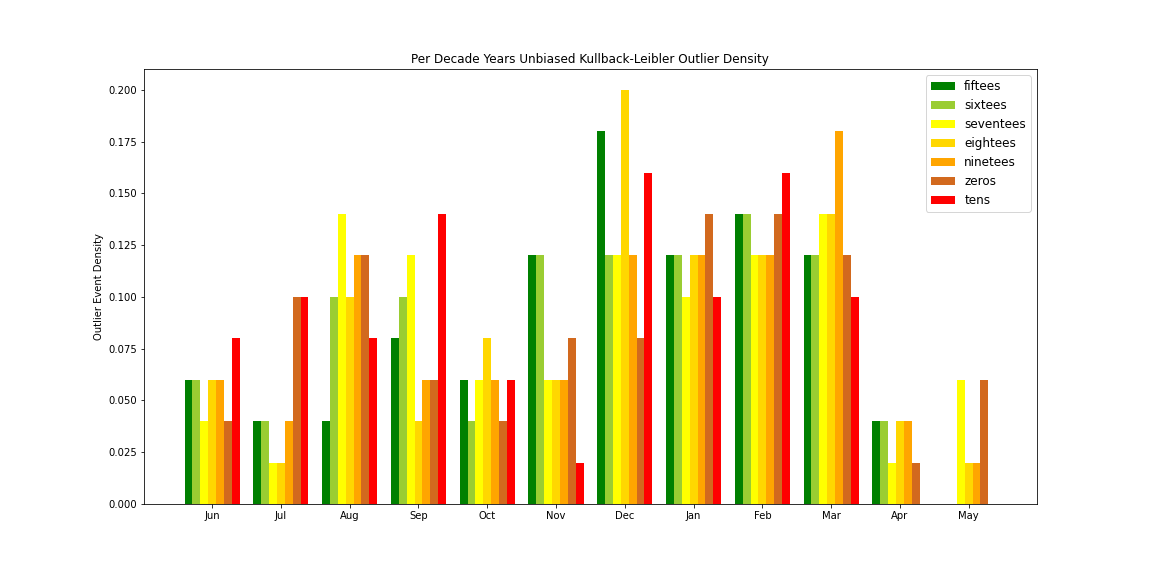}
        \caption{Histogram of the monthly occurrence of outliers per decade. Results for Unbiased Kullback-Leibler divergence measure}
        \label{fig:timing_ts_decades}
\end{figure}

To determine if the timing of outlier events changed throughout the years, the number of events per month of the top 50 outliers is investigated.
These results are depicted in figures \ref{fig:timing_ce_decades} and \ref{fig:timing_ts_decades}, where histograms of the outlier event distribution throughout the months of the year for the Cross Entropy and Unbiased Kullback-Leibler divergence measures are presented.

We found no indication that the month of occurrence changes significantly throughout the years. 
We do note that, as seen in the other experiments, the top outliers detected using Cross Entropy occur in the winter period, while the occurrence of events is more uniform if the Unbiased Kullback-Leibler divergence measure is used. 
The method used thus has significant impact on the type of the detected outlier.

\subsection{Outlier Intensity}
When looking at the univariate total energy generation we see mostly peaks. The number of outliers where the average generation during the outlier, was greater then the average of the overall energy generation is shown in table \ref{tab:outlier_peaktrough}.

We see that troughs in the univariate total energy generation are difficult to detect using the MDI algorithm. Using the Cross Entropy divergence measure only peaks are detected and a potentially risky situation as in Figure 2 remains undetected. This indicates the importance of selecting the correct divergence measure as well as what variables to investigate.
\begin{table}[t]
        \centering
        \caption{Number of peaks and troughs in the top 50 outliers per decade. }
        \label{tab:outlier_peaktrough}
        \begin{tabular}{|l|l|l|l|l|}
                \hline
                Decade & CE Peaks & CE Troughs  & U-KL Peaks  &  U-KL Troughs\\ \hline \hline
                50-60 & 50 & 0 & 41 & 9 \\ \hline
                60-70 & 50 & 0 & 30 & 20 \\ \hline
                70-80 & 50 & 0 & 32 & 18 \\ \hline
                80-90 & 50 & 0 & 37 & 13 \\ \hline
                90-00 & 50 & 0 & 34 & 16 \\ \hline
                00-10 & 50 & 0 & 32 & 18 \\ \hline
                10-20 & 50 & 0 & 29 & 21 \\ \hline
        \end{tabular}
\end{table}

\begin{figure}[t]
        \centering
        \includegraphics[width = \textwidth]{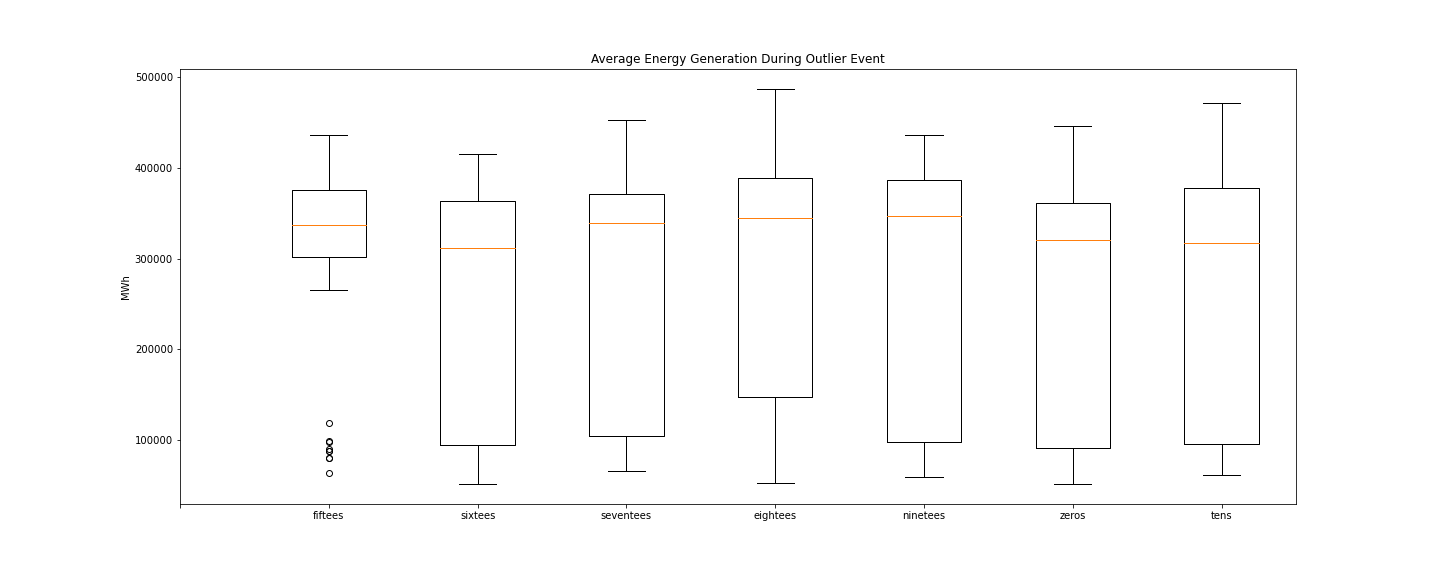}
        \caption{Boxplots of the average hourly Total energy generation during the top 50 outlier events per decade based on the unbiased Kullback-Leibler divergence measure. }
        \label{fig:box_ts_50}
\end{figure}

In the energy generation during an outlier event per decade we observed no periodic behaviour in the top 50 for Unbiased Kullback-Leibler divergence. The effect is hard to see due to the spread of the outlier intensity. It is best observed in figure~\ref{fig:box_ts_50}.

\subsection{Outlier shift and lengthening}
\begin{figure}[ht]
        \centering
        \includegraphics[width = \textwidth]{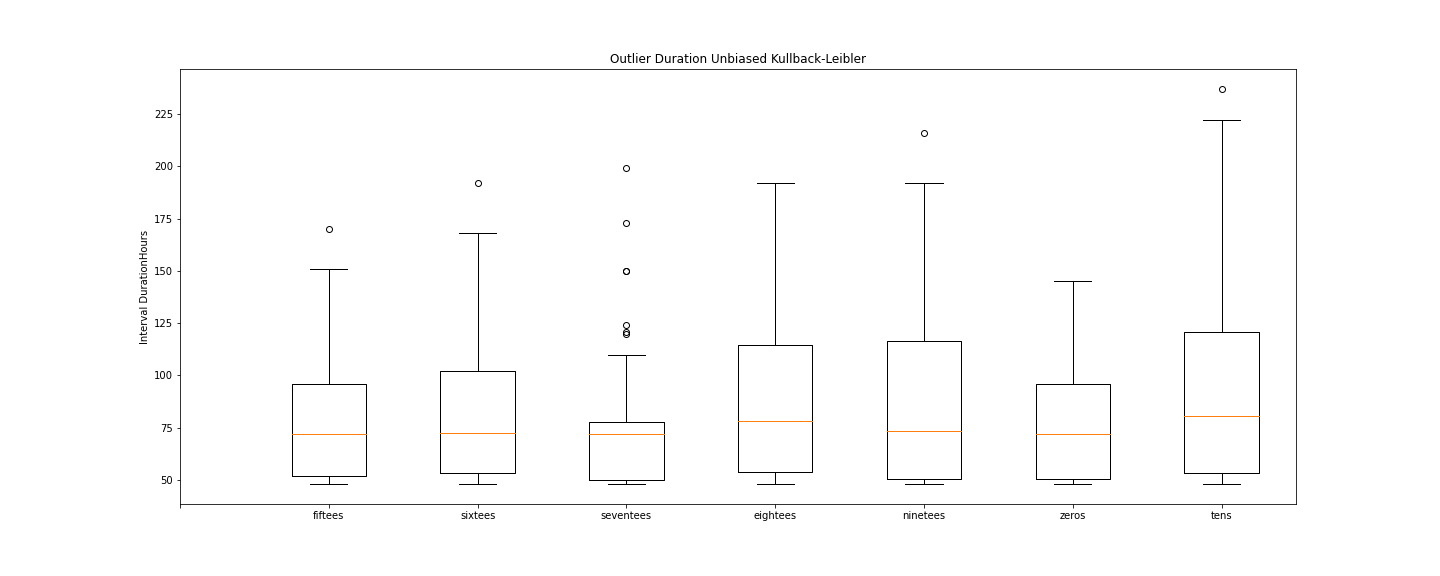}
        \caption{Boxplots of the average interval length in hours per decade. Results for the Cross Entropy divergence measure.}
        \label{fig:interval_ce_decades}
\end{figure}

\begin{figure}[ht]
        \centering
        \includegraphics[width = \textwidth]{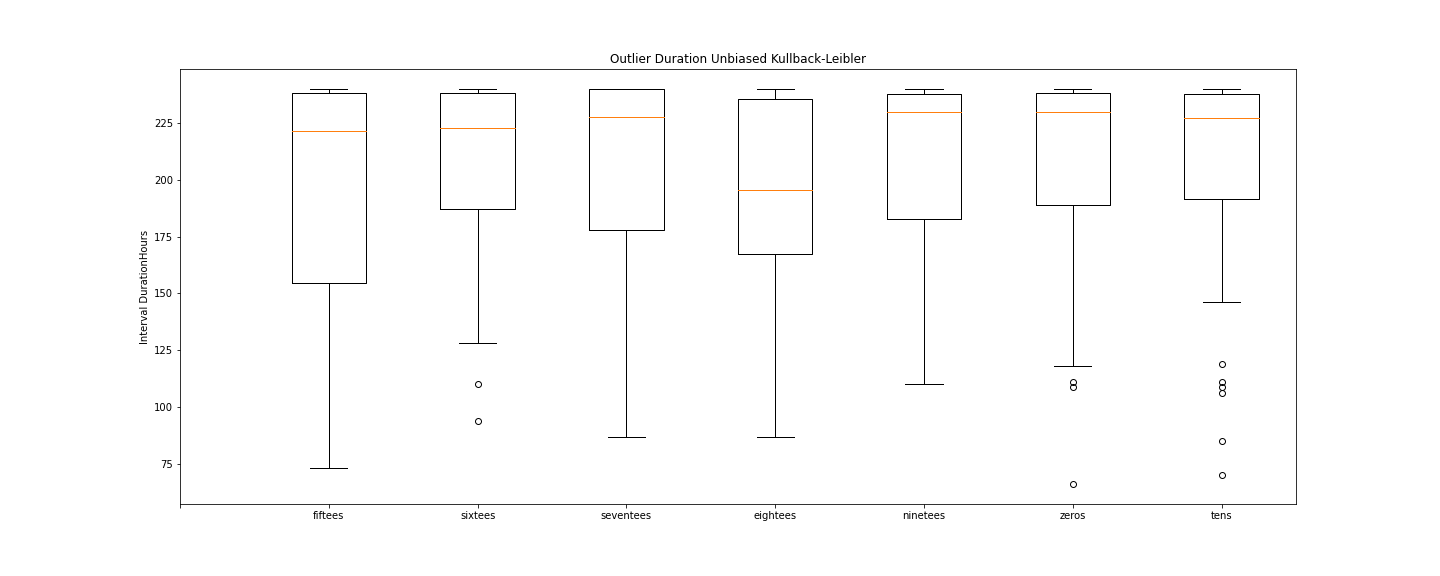}
        \caption{Boxplots of the average interval length in hours per decade. Results for Unbiased Kullback-Leibler divergence measure}
        \label{fig:interval_ts_decades}
\end{figure}

To determine if outlier events are lasting longer throughout the years, the average interval length of the top 50 outliers is investigated.
These results are depicted in figures \ref{fig:interval_ce_decades} and \ref{fig:interval_ts_decades}, where boxplots of the average length for the Cross Entropy and Unbiased Kullback-Leibler divergence measures are presented.

We found no indication that the interval length changes significantly throughout the years. We do note that, as seen in the other experiments, the top outliers detected using Cross Entropy are relatively short and those detected using Unbiased Kullback-Leibler divergence last longer. The method used has significant impact on the length of the detected outlier. This bias might make it more difficult to detect any change in the length of the exact outlying event.

\end{document}